%% file: template.tex
\documentclass{article}

\usepackage{arxiv}

\usepackage[utf8]{inputenc} 
\usepackage[T1]{fontenc}    
\usepackage{hyperref}       
\usepackage{url}            
\usepackage{booktabs}       
\usepackage{amsfonts}       
\usepackage{nicefrac}       
\usepackage{microtype}      
\usepackage{lipsum}
\usepackage{graphicx}
\usepackage{tikz}
\usepackage{subcaption}
\usetikzlibrary{patterns}
\graphicspath{ {./images/} }

\title{A nation-wide experiment: fuel tax cuts and almost free public transport for three months in Germany - Report 1  Study design, recruiting and participation}

\author{
 Allister Loder \\
  School of Engineering and Design\\
  Technical University of Munich\\
  Arcisstrasse 21, 80333 Munich \\
  \texttt{allister.loder@tum.de} \\
   \And
    Fabienne Cantner \\
  TUMCS for Biotechnology \& Sustainability \\ TUM School of Management\\
  Technical University of Munich\\
  Am Essigberg 3, 94315 Straubing \\
  \texttt{fabienne.cantner@tum.de} \\
  \And
  Lennart Adenaw \\
  School of Engineering and Design\\
  Technical University of Munich\\
  Boltzmannstrasse 15, 85748 Garching \\
  \texttt{lennart.adenaw@tum.de} \\
  \And 
  Markus B. Siewert \\
 TUM Think Tank \\
  Munich School of Politics and Public Policy\\
  Richard-Wagner-Straße 1, 80333 München \\
  \texttt{markus.siewert@hfp.tum.de} \\
  \And
  Sebastian Goerg \\
  TUMCS for Biotechnology \& Sustainability \\ TUM School of Management\\
  Technical University of Munich\\
  Am Essigberg 3, 94315 Straubing \\
  \texttt{sebastian.goerg@tum.de} \\
   \And
   Markus Lienkamp \\
  School of Engineering and Design\\
  Technical University of Munich\\
  Boltzmannstrasse 15, 85748 Garching \\
  \texttt{lienkamp@tum.de} \\
  \And
  Klaus Bogenberger \\
   School of Engineering and Design\\
  Technical University of Munich\\
  Arcisstrasse 21, 80333 Munich \\
  \texttt{klaus.bogenberger@tum.de} \\
}

\begin{document}
\maketitle
\begin{abstract}
In spring 2022, the German federal government agreed on a set of measures that aim at reducing households' financial burden resulting from a recent price increase, especially in energy and mobility. These measures include among others, a nation-wide public transport ticket for 9\ EUR per month and a fuel tax cut that reduces fuel prices by more than 15\,\%. In transportation research this is an almost unprecedented behavioral experiment. It allows to study not only behavioral responses in mode choice and induced demand but also to assess the effectiveness of transport policy instruments. We observe this natural experiment with a three-wave survey and an app-based travel diary on a sample of hundreds of participants as well as an analysis of traffic counts. In this first report, we inform about the study design, recruiting and initial participation of study participants.
\end{abstract}


\section{Introduction}

In transportation research, it is quite unlikely to observe or even perform real-world experiments in terms of travel behavior or traffic flow. There are few notable exceptions: subway strikes suddenly make one important alternative mode not available anymore  \cite{Anderson2014,Adler2016}, a global pandemic changes travelers' preferences for traveling at all or traveling collectively with others \cite{Molloy2021}, or a bridge collapse forces travelers to alter their daily activities \cite{Zhu2010}. However, in 2022 the German federal government announced in response to a sharp increase in energy and consumer prices a set of measures that partially offset the cost increases for households. Among these are a public transport ticket at 9\ EUR per month\footnote{\url{https://www.bundesregierung.de/breg-de/aktuelles/9-euro-ticket-2028756}} for traveling all across Germany in public transport, except for long-distance train services (e.g., ICE, TGV, night trains), as well as a tax cut on gasoline and diesel, resulting in a cost reduction of about 15\ \% for car drivers\footnote{\url{https://www.bundesfinanzministerium.de/Content/DE/Standardartikel/Themen/Schlaglichter/Entlastungen/schnelle-spuerbare-entlastungen.html}}. Both measures are limited to three months, namely June, July and August 2022. There are a few noteworthy \textit{planned} large-scale pricing experiments related to this study, e.g., in Copenhagen \cite{Nielsen2008}, in Switzerland \cite{molloy_phd}, and in The Netherlands \cite{PEER2016314}.

In terms of travel behavior, two main effects are expected. First, the changes in travel costs of car and public transport relatively to each other are expected to change travelers' mode choice \cite{ben1985discrete}. Second, the reduction in car traffic and public transport costs is further expected to increase  overall individual mobility, the so-called \textit{rebound effect} \cite{Greening2000,Hymel2010}; this can also be referred to as \textit{induced demand} from a perspective of the generalized costs of travel \cite{Weis2009}. 

For the Munich metropolitan region, Germany, we designed a study comprising three elements. The three elements are: (i) a three-wave survey before, during and after the introduction of cost-saving measures; (ii) a smartphone app based measurement of travel behavior and activities during the same period; (iii) an analysis of aggregated traffic counts and mobility indicators. We will use data from 2019 (pre-COVID-19) and data from shortly before the cost reduction measures as the control group.

\begin{figure}
    \centering
    \includegraphics[width=7cm]{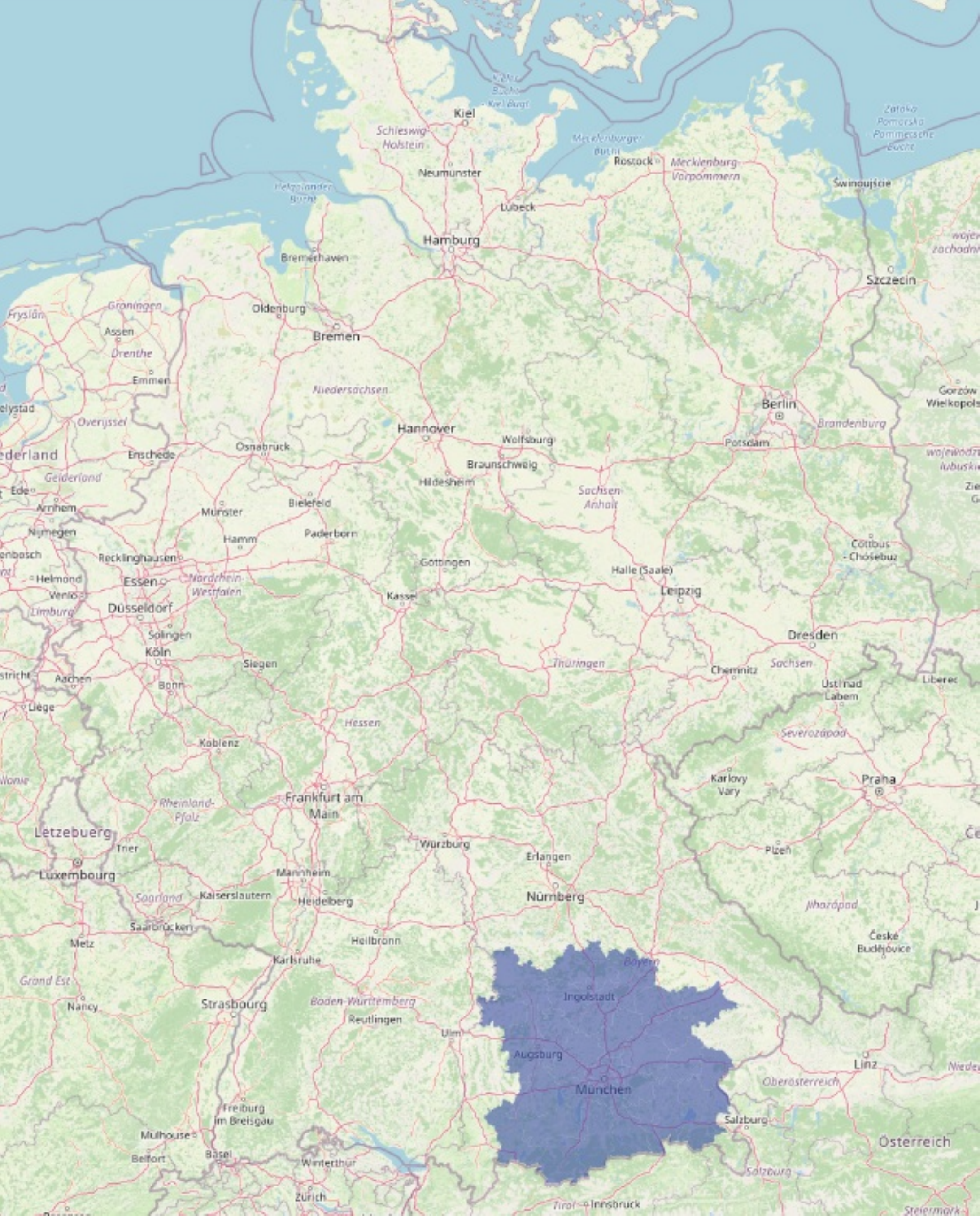}
    \caption{The location of the Munich metropolitan region within Germany. Background map from OpenStreetMap.}
    \label{fig:muc_region}
\end{figure}

The Munich metropolitan region has more than six million inhabitants as is shown in Figure \ref{fig:muc_region}. The region's structure is heterogeneous with a wide range of RegioStaR \footnote{\url{https://www.bmvi.de/SharedDocs/DE/Artikel/G/regionalstatistische-raumtypologie.html}} classification types ranging from rural areas (RegioStaR 77) to highly urbanized metropolitan areas (RegioStaR 71). It comprises three large cities, namely Ingolstadt, Augsburg, and Munich as well as the urbanized areas around and between them. Rural areas to the east and southwest of Munich add to the mix, making the study an ideal representation of different spatial structures and their typical mobility behaviors. Due to the region's size and importance, it has been considered in previous travel surveys and contributed a significant share of participants to the national travel survey Mobilität in Deutschland (MiD) \cite{BundesministeriumfurVerkehrunddigitaleInfrastruktur.2018}. Within the area around Munich, the 2017 MiD reported 3.2 trips per day and person, 42\,km per day and person and 1.5\,h per day and person. In this area, 18\,\% (10\,\%) of all trips are made by public transport and 46\,\% (57\,\%) by car. The numbers in parentheses present German averages\footnote{\url{https://muenchenunterwegs.de/content/657/download/kurzreport-2019.pdf}}.



\section{Study design}

The overall design of the study is shown in Figure \ref{fig:design}. The three main building blocks are a three-wave survey (shown in green), a smartphone-based travel diary (shown in blue), and an analysis of traffic counts and public transport ridership (shown in yellow). In the following, they are introduced separately.

\begin{figure}
    \centering
    \input{_figures/project_orga}
    \caption{Study design of Mobilit\"at.Leben}
    \label{fig:design}
\end{figure}

\subsection{Survey} \label{sec:survey}

The three-wave survey has the following structure: in the first wave, we collect information with regard to mobility tool ownership and we also ask for socio-economic information and socio-political attitudes. The latter include the individual's positioning with respect to the cost reduction measures, climate change and the political environment.  The second survey will include personality questions and further political questions. The concluding third wave will ask respondents again about travel behavior and household spending. In addition, questions about the the cost reduction measures and adaption of new travel and economic behavior as well as future intentions for behavioral change and investments in new technologies will be presented. All three surveys collect information on travel behavior, energy consumption and the effects of inflation on household spending.

The three-wave survey allows us not only to capture the status quo of participants' mobility behavior, but also to measure within-subject changes in behavior and attitudes toward mobility. Additionally, we have the unique possibility to match the stated preferences (and possible changes in these preferences over time) regarding the mobility behavior from the surveys with real behavioral data (revealed preferences). To track the behavioral data we use an app-based travel diary which will be explained in more detail in the next section

\subsection{App-based travel diary} \label{sec:app}

\begin{figure}
    \centering
    \includegraphics[width=4cm]{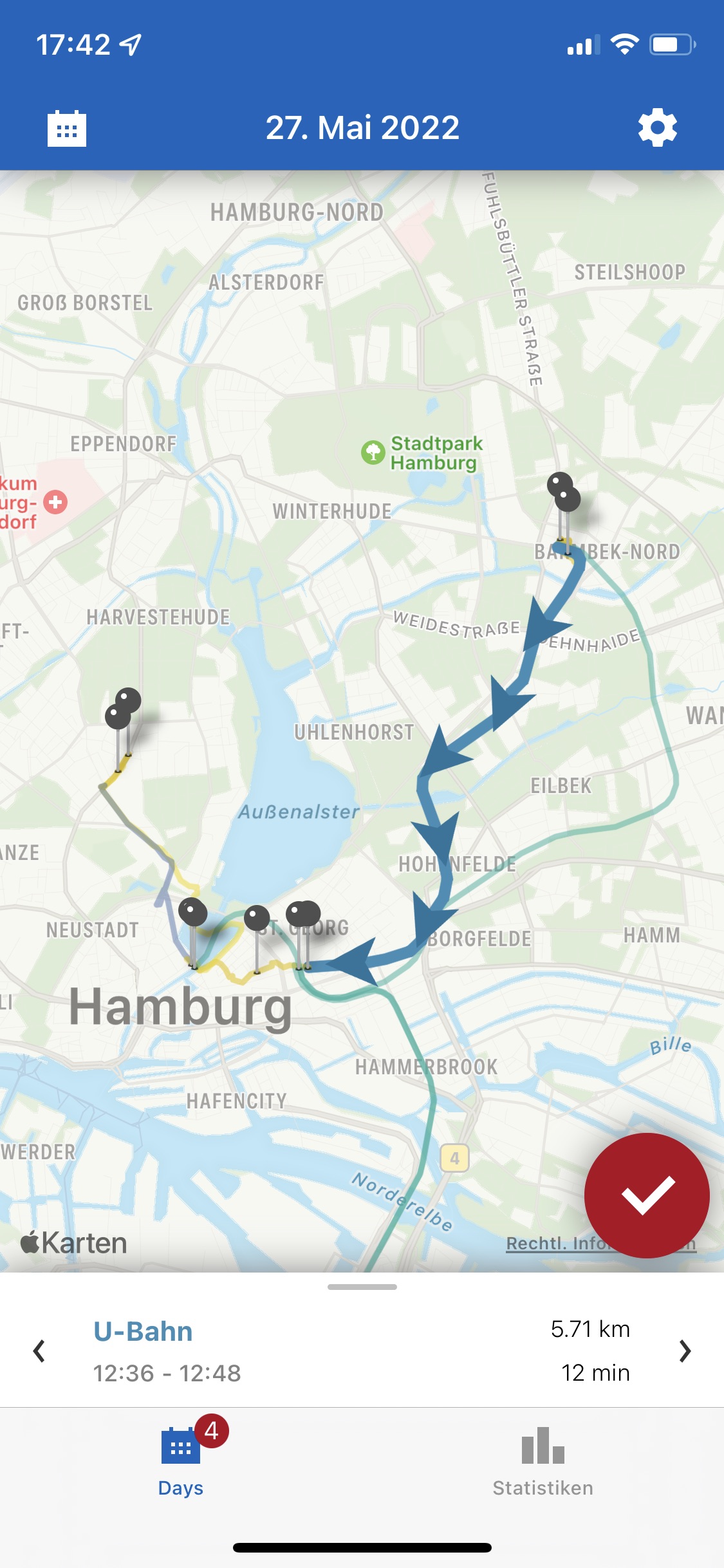}
    \caption{Screenshot of the Mobilit\"at.Leben-App with an example of the collected data.}
    \label{fig:screenapp}
\end{figure}

For our study, we created a dedicated travel diary smartphone app called "Mobilit\"at.Leben". Installed on both iOS and Android devices, it automatically collects participants' waypoints and infers the chosen transport mode as well as the type of activity based on information available from OpenStreetMap. Except for asking participants to occasionally validate their data, there is no need for further interaction with the app. The app records data from the end of May to the end of September 2022. Participants can pause the data collection at any time. Figure \ref{fig:screenapp} shows a screenshot of the developed app. 

The main benefit of using a smartphone app instead of a paper-and-pencil travel diary is that almost every trip gets recorded and measured precisely. 

\subsection{Traffic counts and supplementary data sources} \label{sec:flows}

To corroborate the findings at the individual levels from the survey and the smartphone travel diary, we will rely on aggregated mobility indicators that allow to draw conclusions about traffic volumes and modal shares. Traffic volumes are collected through approximately 6'000 inductive loop detectors spread across the entire city of Munich. These data have successfully shown their capability in explaining the performance of entire road networks \cite{Loder2019SciCap}. Public transport ridership and performance indicators will be made available from the local public transport agency.
Further supplementary data sets relevant to the project (e.g. gas prices and consumer price indices) will be gathered together with regional partners or crawled from public web services in order to allow for a comprehensive analysis of the observed mobility behavior.

These data together with the trip information from the three surveys and travel diaries allow to compute multi-modal city-wide travel production measures \cite{Loder2017}. These measures can then be used, e.g., to assess the effectiveness of cost-reduction policies.

\section{Recruiting strategy}

The recruiting strategy of our study has two parts. First, 1000 participants from all over Germany are approached through a professional panel agency. These participants will only take part in the survey, but not in the smartphone data collection. The rationale of this approach is to obtain an unbiased and representative sample as much as possible. Second, with focus on the Munich metropolitan area, we approach individuals using several media channels, e.g., reports in newspapers, social media, and press conferences. The first sample will serve as a reference for calculating weights for individuals recruited in the second sample. Recruiting for the first sample started on May 25, while the recruiting for the second sample started on May 23.

Participants who complete the three-wave survey as well as collected at least one week of app data per month until September 2022 will receive a reward of 30\ EUR (voucher). Participants who contribute with at least two weeks of data per month will further enter a lottery of three 200\ EUR (voucher). There is no experimental variation in the incentives. 

\section{Participation}

\begin{figure}
     \centering
     \begin{subfigure}[b]{0.45\textwidth}
         \centering
         \includegraphics[width=\textwidth]{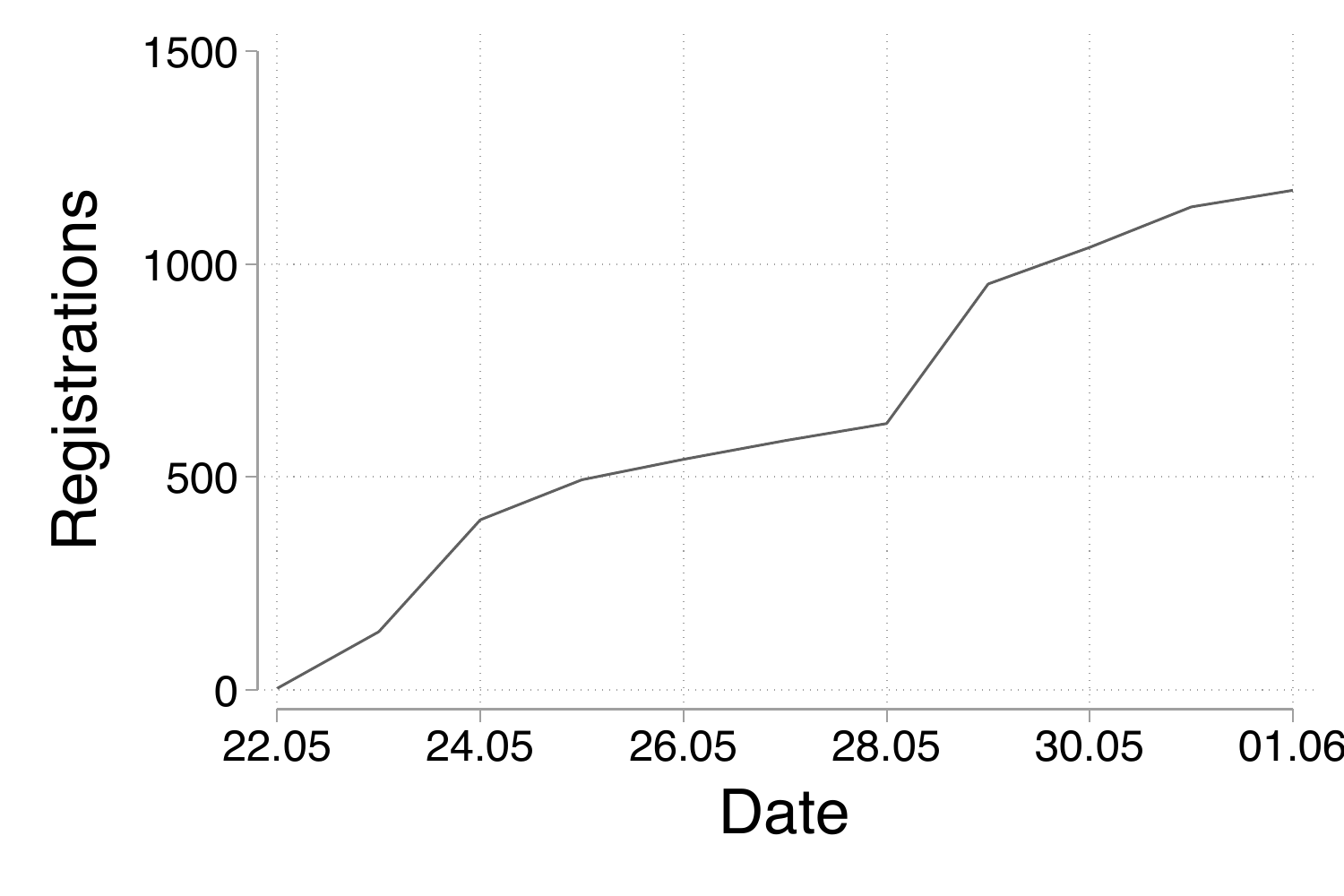}
         \caption{Registrations.}
         \label{fig:registration}
     \end{subfigure}
     \hfill
     \begin{subfigure}[b]{0.45\textwidth}
         \centering
         \includegraphics[width=\textwidth]{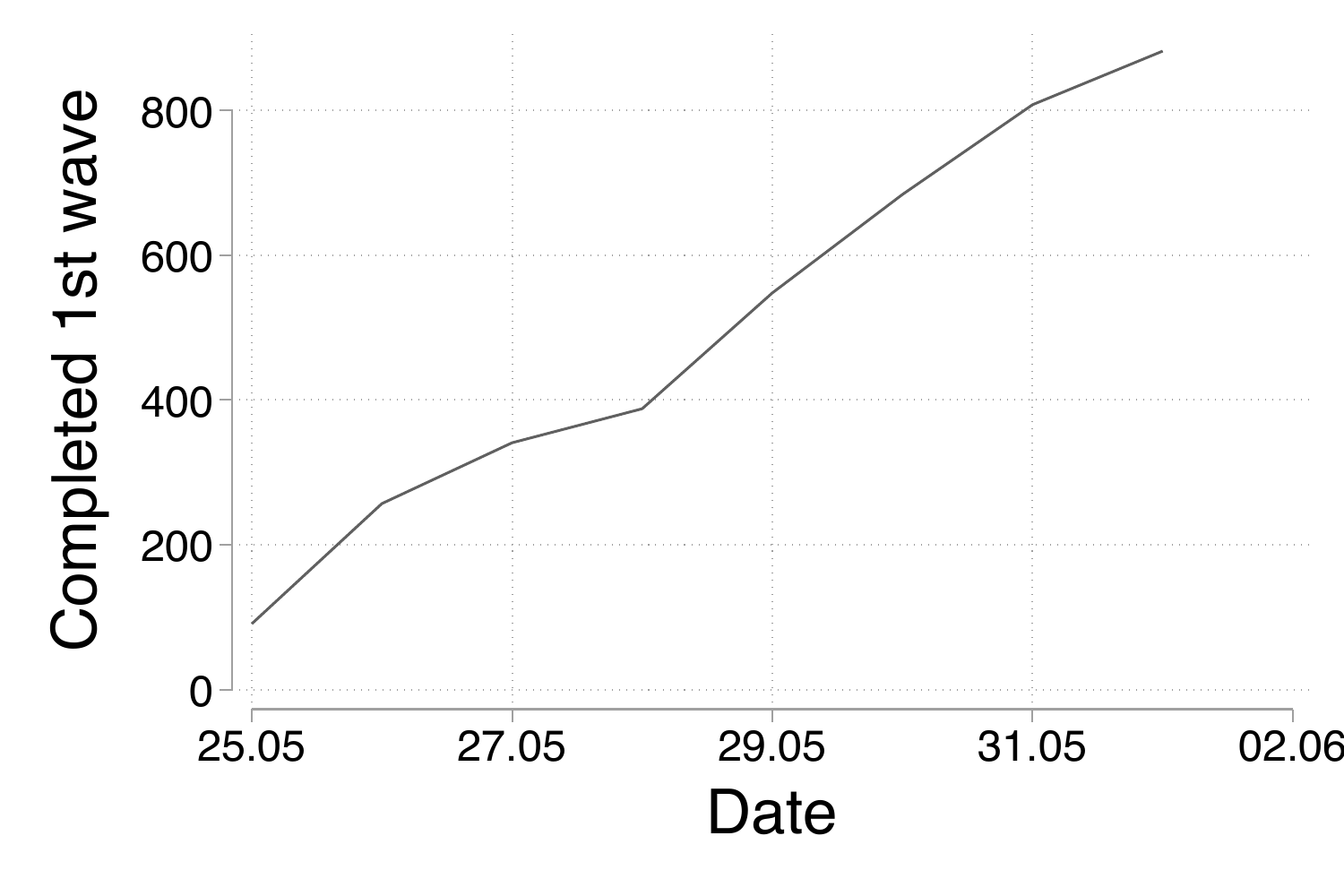}
         \caption{Survey completion.}
         \label{fig:survey}
     \end{subfigure}
     
     \begin{subfigure}[b]{0.45\textwidth}
         \centering
         \includegraphics[width=\textwidth]{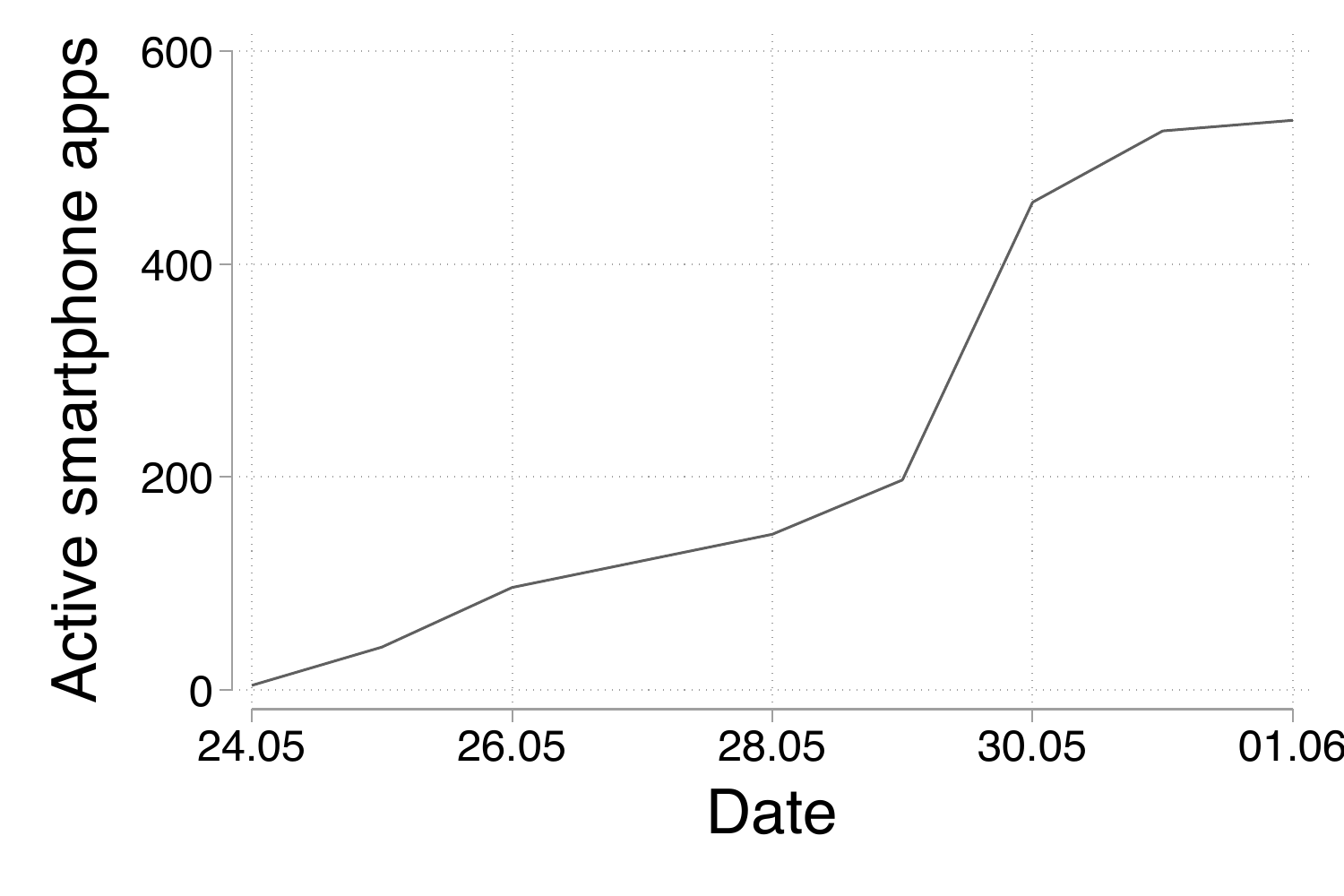}
         \caption{App activation.}
         \label{fig:app}
     \end{subfigure}
        \caption{Recruitment status as of June 1, 2022.}
        \label{fig:involvement}
\end{figure}

Figures \ref{fig:registration} to \ref{fig:app} provide information on the progress of recruiting. In total as of June 1, 2022, 1129 people have successfully registered to participate in our study. 923 want to participate in the survey and the smartphone-study while 206 only want to participate in the survey. Figure \ref{fig:registration} clearly shows the impact of two main events that contributed substantially to increasing participation: a press conference on May 23, 2022 and subsequent media coverage as well as a TV appearance of the study on May 29, 2022. 

So far, more than 80\,\% of invited participants completed the first wave of the survey as seen in Figure \ref{fig:survey}. A reminder has been sent on May 31, 2022 to collect as many responses as possible before the start of the cost reduction measures. 

As seen in Figure \ref{fig:app}, the app activation lags behind the overall registration and survey participation. This results not only from the fact that only 80\,\% of the entire sample decided to use the smartphone app, but in particular as the smartphone apps have been only made available in the app stores on May 25, 2022 (iOS) and May 30, 2022 (Android). Thus, the app activation counts are expected to increase further. Nevertheless, as of June 1, 2022 more than 60\,\% of invited participants are already collecting mobility data.

\begin{figure}
    \centering
    \includegraphics[width=12cm]{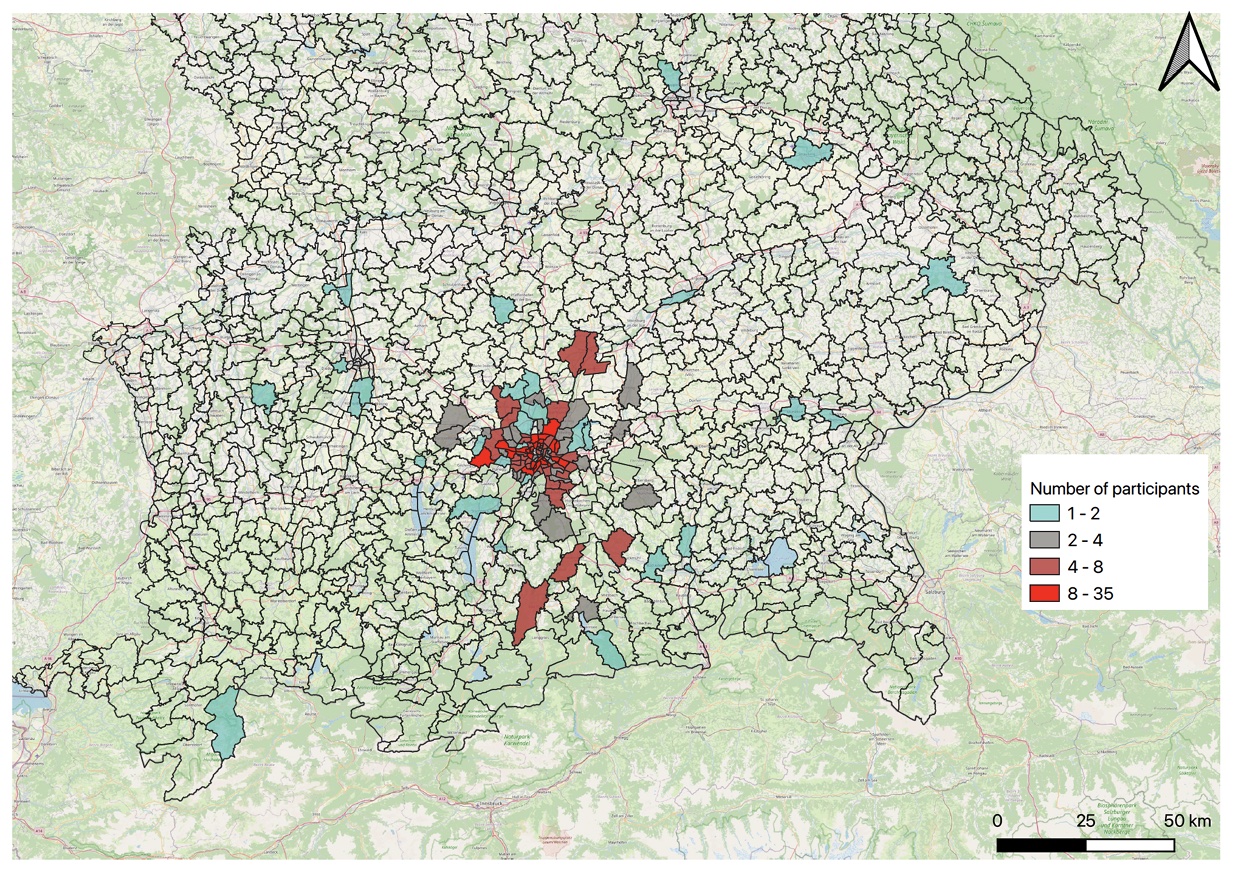}
    \caption{Place of residence of the sample. Background map from OpenStreetMap.}
    \label{fig:home_location}
\end{figure}

\begin{figure}
     \centering
     \begin{subfigure}[b]{0.45\textwidth}
         \centering
         \includegraphics[width=\textwidth]{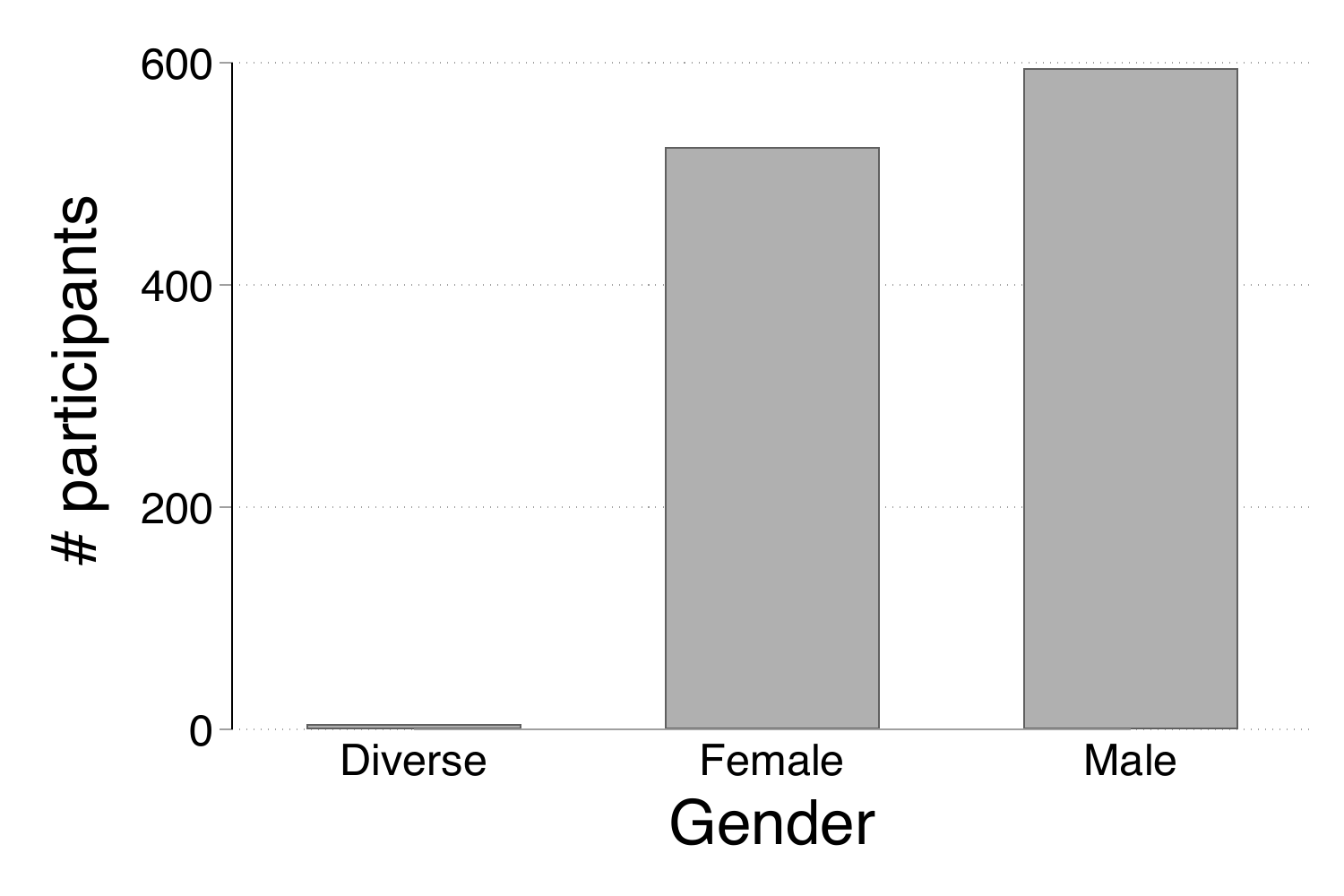}
         \caption{Gender.}
         \label{fig:sex}
     \end{subfigure}
    \hfill
     \begin{subfigure}[b]{0.45\textwidth}
         \centering
         \includegraphics[width=\textwidth]{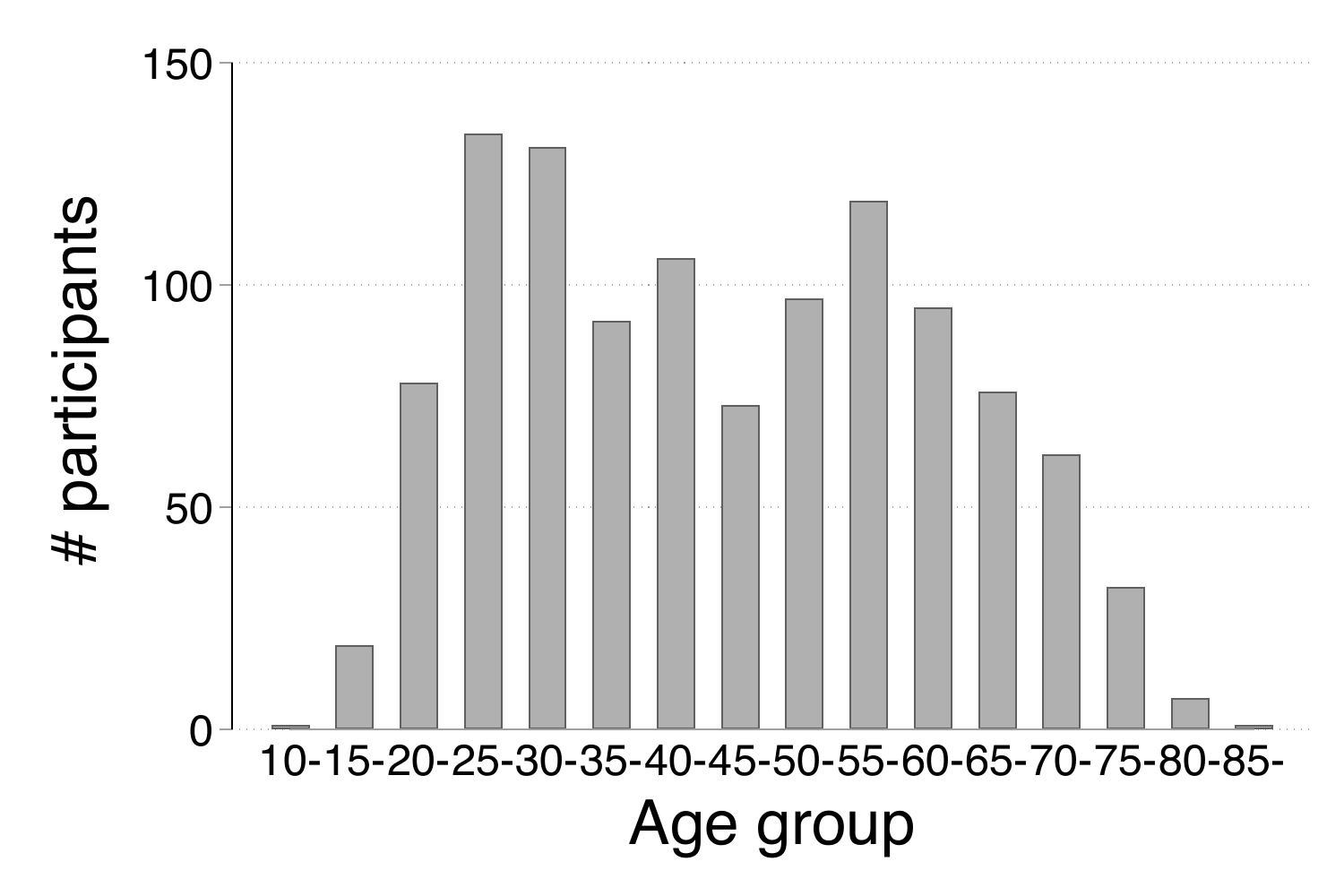}
         \caption{Age.}
         \label{fig:age}
     \end{subfigure}
       
     \begin{subfigure}[b]{0.45\textwidth}
         \centering
         \includegraphics[width=\textwidth]{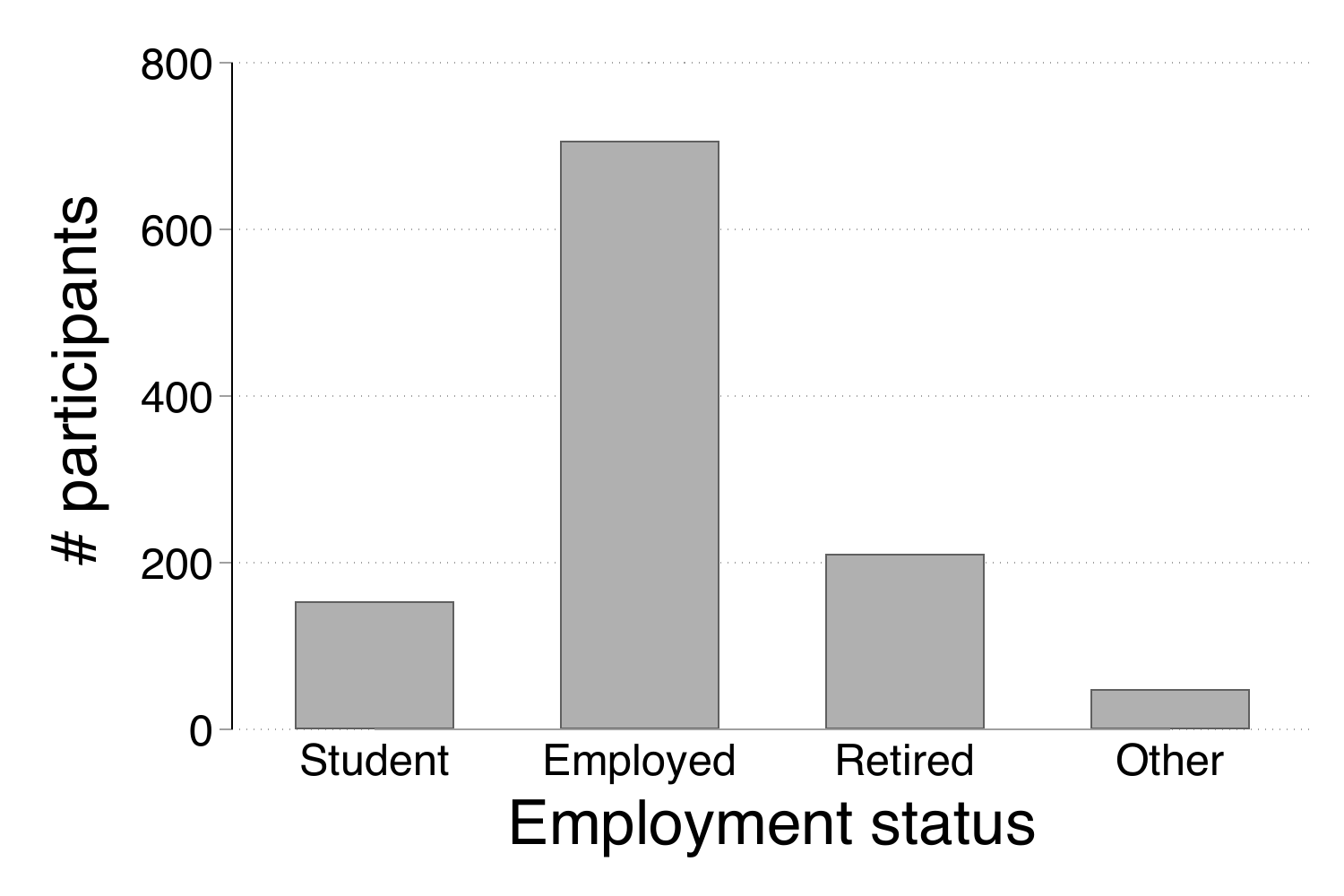}
         \caption{Employment status.}
         \label{fig:employment}
     \end{subfigure}
        \caption{Socio-demographics of the study group.}
        \label{fig:sd}
\end{figure}

Figure \ref{fig:home_location} shows the spatial distribution of the registered participants. The target area of the Munich Metropolitan area is well covered. Figures \ref{fig:sex} to \ref{fig:employment} provide an overview of the core socio-demographic attributes of the participants as of June 1, 2022; one day before the beginning of the cost reduction measures. The participants are almost balanced regarding their gender, with a small bias towards male participants. Age groups are distributed similar to the overall German demographic \cite{BundeszentralefurPolitischeBildung.2021} within the ages 18 to 75. Qualitatively, the participants' age distribution exhibits two local maxima around ages 30 and 55, a local minimum in the age group 45-49 and decreasing proportions of people under 30 and over 60 which is in line with the general population of Germany. In contradiction to the nation-wide age distribution however, the local maximum of the baby-boomers aged 50 to 60 is smaller than the one around the ages of 25 to 35. This will be accounted for during data analysis.

Assuming a population of around 70\,million at the age of 18 and older in Germany, our sample matches the share of employed people pretty well at around 64\,\% \footnote{\url{https://www.destatis.de/DE/Themen/Querschnitt/Statistische-Wochenberichte/wochenberichte-bevoelkerung-xlsx.xlsx;jsessionid=916AE6E504C3DFC634FF3A2191A68469.live712?__blob=publicationFile}}. Contrary, retired people are around five percentage points underrepresented (20 compared to 25\footnote{\url{https://www.destatis.de/DE/Themen/Gesellschaft-Umwelt/Bevoelkerung/Bevoelkerungsstand/_inhalt.html}}.

\section{Outlook}

This is the first report in a sequence of reports planned during the experimental phase to inform academics and the public about the current state of the natural experiment. Once all planned data has been collected and prepared, several quantitative analyses are envisioned: (i) statistical analysis of the main experimental effects based on aggregated variables, e.g., comparing the daily travel distance before, during and after the treatment of cost reductions; (ii) statistical analysis of the individual mode choice determinants; and (iii) statistical analysis of the individual reasons for altering or not altering travel behavior and daily activities as a consequence to the experimental cost changes; (iv) the results from (i) to (iii) will be used to make implications for policy and practice. Further, the project  includes workshops and focus groups with stakeholders to understand more about the behavioral responses in the natural experiment and the their expectations regarding future policy making.

\section*{Acknowledgements}

The authors would like to thank the TUM Think Tank at the Munich School of Politics and Public Policy led by Urs Gasser for their financial and organizational support and the TUM Board of Management for supporting personally the genesis of the project. The authors thank the company MOTIONTAG for their efforts in producing the app at unprecedented speed. Further, the authors would like thank everyone who supported us in recruiting participants, especially Oliver May-Beckmann and Ulrich Meyer from MCube and TUM, respectively.

\bibliographystyle{unsrt}  
\bibliography{references}  



\end{document}

%% file: _figures/project_orga.tex
 
\tikzset{
pattern size/.store in=\mcSize, 
pattern size = 5pt,
pattern thickness/.store in=\mcThickness, 
pattern thickness = 0.3pt,
pattern radius/.store in=\mcRadius, 
pattern radius = 1pt}
\makeatletter
\pgfutil@ifundefined{pgf@pattern@name@_keincrx0y}{
\pgfdeclarepatternformonly[\mcThickness,\mcSize]{_keincrx0y}
{\pgfqpoint{0pt}{0pt}}
{\pgfpoint{\mcSize+\mcThickness}{\mcSize+\mcThickness}}
{\pgfpoint{\mcSize}{\mcSize}}
{
\pgfsetcolor{\tikz@pattern@color}
\pgfsetlinewidth{\mcThickness}
\pgfpathmoveto{\pgfqpoint{0pt}{0pt}}
\pgfpathlineto{\pgfpoint{\mcSize+\mcThickness}{\mcSize+\mcThickness}}
\pgfusepath{stroke}
}}
\makeatother
\tikzset{every picture/.style={line width=0.75pt}} 

\begin{tikzpicture}[x=0.75pt,y=0.75pt,yscale=-1,xscale=1]

\draw  [fill={rgb, 255:red, 208; green, 2; blue, 27 }  ,fill opacity=0.64 ] (175,101) -- (360,101) -- (360,270) -- (175,270) -- cycle ;
\draw  [fill={rgb, 255:red, 255; green, 255; blue, 255 }  ,fill opacity=1 ] (116.5,109) -- (172.5,109) -- (172.5,133) -- (116.5,133) -- cycle ;
\draw  [fill={rgb, 255:red, 255; green, 255; blue, 255 }  ,fill opacity=1 ] (177.5,109) -- (233.5,109) -- (233.5,133) -- (177.5,133) -- cycle ;
\draw  [fill={rgb, 255:red, 255; green, 255; blue, 255 }  ,fill opacity=1 ] (239.5,109) -- (295.5,109) -- (295.5,133) -- (239.5,133) -- cycle ;
\draw  [fill={rgb, 255:red, 255; green, 255; blue, 255 }  ,fill opacity=1 ] (300.5,109) -- (356.5,109) -- (356.5,133) -- (300.5,133) -- cycle ;
\draw  [fill={rgb, 255:red, 255; green, 255; blue, 255 }  ,fill opacity=1 ] (362.5,109) -- (418.5,109) -- (418.5,133) -- (362.5,133) -- cycle ;
\draw  [fill={rgb, 255:red, 255; green, 255; blue, 255 }  ,fill opacity=1 ] (425.5,109) -- (481.5,109) -- (481.5,133) -- (425.5,133) -- cycle ;
\draw  [fill={rgb, 255:red, 255; green, 255; blue, 255 }  ,fill opacity=1 ] (488.5,109) -- (544.5,109) -- (544.5,133) -- (488.5,133) -- cycle ;
\draw  [fill={rgb, 255:red, 74; green, 144; blue, 226 }  ,fill opacity=1 ] (158,139) -- (411.5,139) -- (411.5,163) -- (158,163) -- cycle ;
\draw  [fill={rgb, 255:red, 184; green, 233; blue, 134 }  ,fill opacity=1 ] (164.29,108.45) -- (148.45,88.03) -- (179.96,88) -- cycle ;
\draw  [fill={rgb, 255:red, 184; green, 233; blue, 134 }  ,fill opacity=1 ] (286.29,108.45) -- (270.45,88.03) -- (301.96,88) -- cycle ;
\draw  [fill={rgb, 255:red, 184; green, 233; blue, 134 }  ,fill opacity=1 ] (403.29,108.96) -- (387.45,88.54) -- (418.96,88.51) -- cycle ;
\draw  [fill={rgb, 255:red, 248; green, 231; blue, 28 }  ,fill opacity=1 ] (8,169) -- (518,169) -- (518,196) -- (8,196) -- cycle ;
\draw  [fill={rgb, 255:red, 255; green, 255; blue, 255 }  ,fill opacity=1 ] (8,110) -- (64,110) -- (64,134) -- (8,134) -- cycle ;
\draw  [pattern=_keincrx0y,pattern size=6pt,pattern thickness=0.75pt,pattern radius=0pt, pattern color={rgb, 255:red, 74; green, 144; blue, 226}] (9,139) -- (64,139) -- (64,163) -- (9,163) -- cycle ;
\draw   (4,6) -- (69,6) -- (69,279) -- (4,279) -- cycle ;
\draw   (109,6) -- (552,6) -- (552,279) -- (109,279) -- cycle ;

\draw (129,112.01) node [anchor=north west][inner sep=0.75pt]  [rotate=-359.98] [align=left] {May};
\draw (190,112.01) node [anchor=north west][inner sep=0.75pt]   [align=left] {June};
\draw (252,112.01) node [anchor=north west][inner sep=0.75pt]   [align=left] {July};
\draw (313,112.01) node [anchor=north west][inner sep=0.75pt]   [align=left] {Aug.};
\draw (375,112.01) node [anchor=north west][inner sep=0.75pt]   [align=left] {Sept};
\draw (438,112.01) node [anchor=north west][inner sep=0.75pt]   [align=left] {Oct.};
\draw (500,112.01) node [anchor=north west][inner sep=0.75pt]   [align=left] {Nov.};
\draw (208,143) node [anchor=north west][inner sep=0.75pt]   [align=left] {Smartphone travel diary};
\draw (141,43) node [anchor=north west][inner sep=0.75pt]   [align=left] {\begin{minipage}[lt]{37.3pt}\setlength\topsep{0pt}
\begin{center}
Survey \\wave 1
\end{center}

\end{minipage}};
\draw (262,43) node [anchor=north west][inner sep=0.75pt]   [align=left] {\begin{minipage}[lt]{37.87pt}\setlength\topsep{0pt}
\begin{center}
Survey\\ wave 2
\end{center}

\end{minipage}};
\draw (378,43) node [anchor=north west][inner sep=0.75pt]   [align=left] {\begin{minipage}[lt]{37.87pt}\setlength\topsep{0pt}
\begin{center}
Survey\\ wave 3
\end{center}

\end{minipage}};
\draw (114,174) node [anchor=north west][inner sep=0.75pt]   [align=left] {Traffic counts and public transport ridership};
\draw (184,225) node [anchor=north west][inner sep=0.75pt]   [align=left] {\begin{minipage}[lt]{116.67pt}\setlength\topsep{0pt}
\begin{center}
Period of \\ cost reduction measures
\end{center}

\end{minipage}};
\draw (19,11.01) node [anchor=north west][inner sep=0.75pt]  [rotate=-359.98] [align=left] {2019};
\draw (9,292) node [anchor=north west][inner sep=0.75pt]   [align=left] {{\scriptsize *) Travel diaries from 2019 are available from another sample. }};
\draw (32,142) node [anchor=north west][inner sep=0.75pt]   [align=left] {*)};
\draw (314,12) node [anchor=north west][inner sep=0.75pt]  [rotate=-359.98] [align=left] {2022};
\draw (15,114) node [anchor=north west][inner sep=0.75pt]  [rotate=-359.98] [align=left] {Spring};
\draw (81.5,11) node [anchor=north west][inner sep=0.75pt]   [align=left] {...};

\end{tikzpicture}